\documentclass[twocolumn,amsmath,amssymb,amsfonts]{revtex4}
\usepackage{graphicx}
\usepackage{amsmath}

\newcommand{\eps}{\varepsilon}

\newcommand{\sgn}{\mathrm{sgn}}
\newcommand{\beq}{\begin{equation}}
\newcommand{\eeq}{\end{equation}}

\def\barr{\begin{array}}
\def\earr{\end{array}}
\def\half{{\textstyle {1\over 2}}}
\def\third{{\textstyle {1\over 3}}}

\begin{document}

% Front matter
\title{Expansion shock waves in regularised shallow water theory}

\author{Gennady A.~El}
\affiliation{Department of Mathematical Sciences,
    Loughborough University, Loughborough, LE11 3TU, United Kingdom}
\author{Mark A.~Hoefer}
\affiliation{Department of Applied Mathematics, University of
  Colorado, Boulder, CO, 80309, USA}
\author{Michael Shearer}
\affiliation{Department of Mathematics, North Carolina State
  University, Raleigh, NC, 27695, USA}

% Abstract

\begin{abstract}
 We identify a new type of shock wave by constructing a stationary
  expansion shock solution of a class of regularised shallow water
  equations that include the Benjamin-Bona-Mahoney (BBM) and
  Boussinesq equations. An expansion shock exhibits divergent
  characteristics, thereby contravening the classical Lax entropy condition.
  The persistence of the expansion shock in initial value problems is
  analysed and justified using matched asymptotic expansions and
  numerical simulations.  The expansion shock's existence is traced to
  the presence of a non-local dispersive term in the governing
  equation. We establish the algebraic decay of the shock as it is
  gradually eroded by a simple wave on either side. More generally, we
  observe a robustness of the expansion shock in the presence of weak
  dissipation and in simulations of asymmetric initial conditions
  where a train of solitary waves is shed from one side of the shock.
\end{abstract}


\maketitle

\section{Introduction}
\label{sec:introduction}

In this paper, we consider a class of nonlinear, dispersive equations
naturally arising in shallow water theory, most concisely exemplified
by a version of the Benjamin-Bona-Mahoney (BBM) equation, also known
as the regularised long wave equation
\begin{equation}
  \label{eq:1}
  u_t +uu_x =   u_{xxt}.
\end{equation}
The original BBM equation, which contains an additional linear
convective term $u_x$, is an important model for the description of
unidirectional propagation of weakly nonlinear, long waves in the
presence of dispersion. It first appeared in a numerical study of
shallow water undular bores, \cite{peregrine_calculations_1966}, and
later was proposed in \cite{benjamin_model_1972} as an analytically
advantageous alternative to the Korteweg-de Vries (KdV) equation
\begin{equation}
  \label{kdv1}
  u_t + uu_x =  -u_{xxx}.
\end{equation}

In the context of shallow water waves, the BBM and KdV equations
(\ref{eq:1}) and (\ref{kdv1}) are reduced, normalised versions of
corresponding asymptotic models derived from the general Euler
equations of fluid mechanics using small amplitude, long wave
expansions. If $\delta \ll 1$ is the ratio of the undisturbed depth to
a typical wave length and $\epsilon \ll1$ is the ratio of a typical
wave amplitude to the undisturbed depth, then the asymptotic KdV and
BBM equations occur under the balance $\epsilon \sim \delta^2,$
\cite{whitham_linear_1974}, and so can be used interchangeably within
their common domain of asymptotic validity, \cite{johnson2002}.

Despite asymptotic equivalence, the mathematical properties of the BBM
and KdV equations are very different, which is acutely captured by
their normalised versions (\ref{eq:1}) and (\ref{kdv1}). The KdV
equation (\ref{kdv1}) is known to be integrable via the inverse
scattering transform and to possess an infinite number of conservation
laws. The BBM equation (\ref{eq:1}), by contrast, does not enjoy full
integrability and has only three independent conservation laws.
Nonetheless, well-posedness of initial value problems for both
equations has been established in the Sobolev spaces $H^s$ (with $s>0$
for BBM \cite{bona_tzvetkov_2009}, $s>-\frac 34$ for KdV,
\cite{colliander_2003}).

As a numerical and mathematical model, the BBM equation yields more satisfactory
short-wave behaviour,  due to the
regularisation of the unbounded growth in frequency, phase and group
velocity values present in the KdV equation.  In particular, this enables less strict time-stepping in numerical schemes for the BBM equation. Indeed, linearising
(\ref{eq:1}) about a constant $u=u_0$: $u(x,t)=u_0+ae^{i(kx-\omega
  t)}$, we obtain the dispersion relation
\begin{equation}
  \label{dispersion}
  \omega=\omega_0(k;u_0)=u_0\frac{k}{1+ k^2}.
\end{equation}
The phase and group velocities are 
\begin{equation}
  \label{gp_velocity}
  c_p= \frac{\omega_0}{k}= \frac{u_0}{1+ k^2} , \quad c_g=\partial_k \omega_0
  = u_0\frac{1- k^2}{(1+ k^2)^2}.
\end{equation}
One can see that $\omega_0$ as well as $c_p$ and $c_g$ are bounded as
functions of the wave number $k$, in contrast to their counterparts
for the KdV equation with dispersion relation $\omega_0=u_0 k-k^3$.
The rational form of BBM dispersion \eqref{dispersion} indicates its
non-local character.  Moreover, the dynamics of \textit{linear}
dispersive equations with discontinuous initial data exhibit distinct
qualitative structure depending upon bounded or unbounded dispersion
behaviour for large $k$ \cite{chen_olver_2012}.

We remark that ``engineering'' the dispersive properties of model
equations was pioneered by Whitham in the context of water waves, see
\cite{whitham_variational_1976}.  In addition to some already
mentioned mathematical and numerical advantages, one may also achieve
superior physical accuracy, when compared with standard asymptotic
models, by incorporating full linear dispersion,
\cite{moldabayev_whitham_2015}.

Equations (\ref{eq:1}) and (\ref{kdv1}) represent two different
dispersive mechanisms to regularise the scalar conservation law, the
inviscid Burgers equation
\begin{equation}
  \label{inviscid_burg} 
  u_t+ (\frac{1}{2} u^2)_x=0.  
\end{equation}
Dispersive regularisation of hyperbolic conservation laws is known to
give rise to dispersive shock waves (DSWs), also known as undular
bores, \cite{gurevich_nonstationary_1974,fornberg_numerical_1978},
which are in many respects very different from their diffusive or
diffusive-dispersive counterparts, \cite{el_dispersive_2015}. These
DSWs have a distinct oscillatory structure and expand with time so
that the Rankine-Hugoniot relations are not applicable to
them. Instead, DSW closure is achieved via an appropriate solution of
the Whitham modulation equations obtained by a nonlinear wave
averaging procedure applied to the full dispersive equation,
\cite{whitham_linear_1974,el_dsws_2016}.  Dispersive shock waves are
evolutionary if they satisfy causality conditions, \cite{el_dsws_2016}
and thus represent dispersive counterparts of classical, Lax shocks,
\cite{lax1957}.  All shock solutions of the KdV equation are
evolutionary DSWs, \cite{el_dsws_2016}. In contrast, we show in this
paper that the BBM equation (\ref{eq:1}) admits a family of stationary
(non-propagating), non-oscillatory {\it expansion shocks} that (i)
satisfy the Rankine-Hugoniot jump condition, and (ii) violate
causality. BBM expansion shocks are very different from both classical
shocks of the inviscid Burger's equation (\ref{inviscid_burg}) and
DSWs of the KdV equation (\ref{kdv1}).

Nonlinear partial differential equations of hyperbolic type, such as
those modelling inviscid gas dynamics, e.g., (\ref{inviscid_burg}),
can have discontinuous solutions. These weak solutions may or may not
be physical, depending on whether they are stable, or persist under
small changes to initial conditions or the governing equations.  Shock
waves are physical, discontinuous solutions that typically satisfy
side conditions associated with either a physical or mathematical
notion of entropy. In gas dynamics, these conditions force shock waves
to be compressive in that they compress the gas as they pass a fixed
location. This kind of condition was expressed by Lax in the 1950s in
terms of characteristics, requiring that shock waves are evolutionary,
i.e., they are uniquely determined from initial conditions.

In this paper, we show that a non-evolutionary stationary shock wave
of the BBM equation (\ref{eq:1}) persists but decays algebraically in
time. This example is surprising because hyperbolic theory would
suggest that the stationary shock would immediately give way to a
continuous solution, namely a rarefaction wave. The persistence is
explained through the interaction of the particular non-local nature
of dispersion in the BBM equation and a length scale associated with
the stationary shock, that sets the time scale for decay.  
% Incidentally, discontinuous periodic waves were studied for a
% variety of dispersive equations in \cite{chen_olver_2012}.

Expansion shocks are not unique to the BBM equation.  We show that
they also persist in one of the versions of the classical
bi-directional Boussinesq equations for dispersive shallow-water
waves, \cite{boussinesq, whitham_linear_1974}.  Similar to the BBM
equation, these Boussinesq equations have the term $u_{xxt}$ in the
momentum equation. (Existence of weak solutions of initial value
problems for Boussinesq equations was established in
\cite{schonbek_1981}.) More broadly, we identify a large class of
non-evolutionary partial differential equations --- i.e., equations
not explicitly resolvable with respect to the first time derivative,
\cite{SAPM:SAPM376} --- that exhibit decaying expansion shock
solutions, indicating the ubiquity of these new solutions.

\section{Shocks and rarefactions}

If the dispersive right hand side of the BBM (\ref{eq:1}) or KdV
(\ref{kdv1}) equation is deleted, we are left with the inviscid
Burger's equation (\ref{inviscid_burg}), a scalar conservation law
that admits shock wave weak solutions 
\begin{equation}
  \label{shock}
  u(x,t)=\left\{\begin{array}{ll}
      u_-, \quad & x<st\\[6pt]
      u_+, &x>st,
    \end{array}
  \right.
\end{equation}
provided the speed $s$ is the average of the characteristic speeds
$u_\pm$ on either side of the shock: $s=\half(u_++u_-)$.  Such shocks
are stable %\cite{serre}
provided that characteristics enter the shock from each side,
$u_+<u_-$, a condition known as the Lax entropy condition
\cite{lax1957}. In this case, the shock is called an entropy shock, or
by analogy with gas dynamics, a {\em compressive} shock.

By contrast, a shock wave (\ref{shock}) solution of
(\ref{inviscid_burg}) is called {\em expansive} if $u_-<u_+.$
Expansion shocks are thought to be unstable and to violate causality,
because characteristics leave rather than approach the shock.  Instead
of an expansion shock, a self-similar rarefaction wave resolves the
discontinuity between $u_-$ and $u_+$:
\begin{equation}
  \label{raref1}
  u(x,t)=\left\{\begin{array}{ll}
      u_-, \quad & x<u_-t\\[6pt]
      x/t, & u_-t<x<u_+t\\[6pt]
      u_+, &x>u_+t.
    \end{array}
  \right.
\end{equation}
 
When $u_+=-u_-,$ the shock wave (\ref{shock}) is stationary, and hence
is also a weak solution of the BBM equation (\ref{eq:1}) (since the
shock is time-independent). With $u_->0$, the stable case, the
stationary shock persists.  However, for $u_-<0,$ the unstable case,
hyperbolic theory would suggest that the jump is immediately replaced
by the self-similar rarefaction wave (\ref{raref1}) or some
approximation to it. However, we find that the dispersive
regularisation resulting from the BBM equation sustains solutions in
which a smoothed stationary shock persists but decays algebraically,
as shown in Fig.~\ref{fig:expansion_shock_sequence}. More precisely,
we study the initial value problem with initial data being a smoothed
stationary shock, with width $\eps>0$.

\section{The expansion shock.}

\begin{figure}  
    \centering
   \includegraphics[width=\columnwidth]{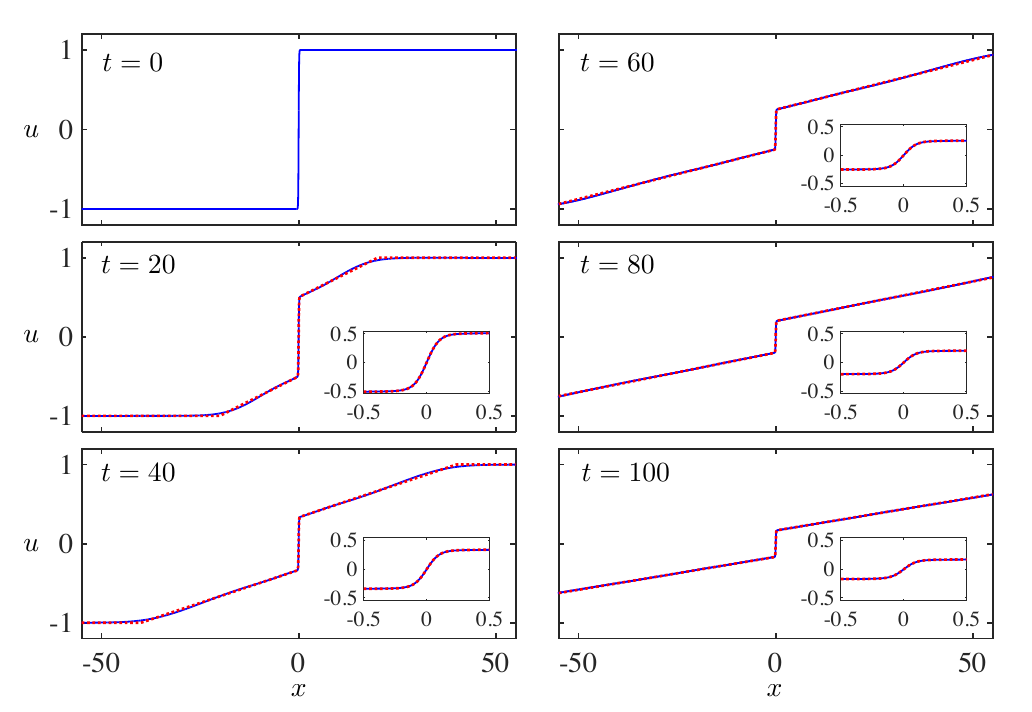}
   \caption{Numerical (solid, blue) and asymptotic (dashed, red)
     solutions of the initial value problem for eq.~(\ref{eq:1}) with
     initial data eq.~(\ref{ic1}) where $\eps = 0.1$ and $A = 1$.
     \label{fig:expansion_shock_sequence}}
 \end{figure}

To see the effect of dispersion on a stationary shock, we pose initial
data
\begin{equation}
  \label{ic1}
  u(x,0)=A\tanh\frac x\eps, \quad -\infty<x<\infty,
\end{equation}
with amplitude $A>0$ for the BBM equation (\ref{eq:1}). Thus, as
$\eps\to 0$, the initial data converge to a jump from $u=-A$ to $u=A,$
representing a stationary expansion shock solution to the inviscid
Burger's equation (\ref{inviscid_burg}). The numerical solution of
(\ref{eq:1}), (\ref{ic1}) is shown in
Fig.~\ref{fig:expansion_shock_sequence}.
% We use a pseudospectral spatial discretization with fourth order
% Runge-Kutta time-stepping, similar to the method described in
% \cite{el_dispersive_2015}.  The numerical computations were
% performed on the domain $x \in [-L,L]$ with $N$ Fourier modes and
% the timestep $\Delta t$.  See Figure captions for parameter values.
We observe the development of a rarefaction wave on either side of a
stationary but decaying shock. We analyse the solution
by matched asymptotics using $\eps$ as the small parameter. First note
that the initial function $u(x,0)$ is an odd function, and the
solution $u(x,t)$ should therefore be an odd function of $x$ for each
$t>0.$

\subsection{The inner solution} 
To capture the inner solution, we introduce  into eq.~(\ref{eq:1}) the short space $\xi =
x/\eps$ and long time $T=\eps t$ scalings of the independent variables
$x$ and $t$
\begin{equation}
  \label{asymptotics2} 
  \eps u_T+\frac 1\eps u u_\xi = \frac 1\eps
   u_{\xi\xi T}.
\end{equation}
Expanding the dependent variable
\begin{equation}
  \label{asymptotics1}
  u=u_0(\xi,T)+\eps u_1(\xi,T) +\dots,
\end{equation}
and substituting this ansatz into (\ref{asymptotics2}) yields
the leading order equation
\begin{equation*}
  u_0 \partial_\xi u_0= \partial_{\xi\xi T}u_0.
\end{equation*}
This equation admits separated solutions $u_0(\xi,T)=f(\xi)a(T)$,
leading to
$$
ff' a^2=f'' \dot{a}, 
$$
where $ '$, $\dot{\ }$\ denote derivatives with respect to $\xi$ and $T$,
respectively.  Introducing a separation constant $K > 0$ 
$$
\frac{\dot{a}}{a^2}=\frac{ff'}{f''}=-K,
$$
we obtain the solution
$$
a(T)=\frac{a_0}{1+a_0KT}, 
$$
and 
$$
f(\xi)=\sqrt{c}\tanh\left(\frac{\sqrt{c}}{2K}(\xi-\xi_0)\right).  
$$
In these formulas, $K,a_0, c$ and $\xi_0$ are arbitrary constants. To
agree with the initial data (\ref{ic1}), we set $c=1, K=\frac 12,
a_0=A,$ and $\xi_0=0.$ Thus, the leading order inner solution is
\begin{equation}
  \label{inner}
  u \sim u_{in}(\xi,T)=\frac{A}{1+\half A T} \tanh \xi .
\end{equation}
The inner solution reveals the smoothed structure of the dispersively
regularised expansion shock and its algebraic temporal decay.

\subsection{The outer solution} 
The outer solution has a different, long space and time scaling
\begin{equation*}
  X=\eps x, \quad T=\eps t.
\end{equation*}
This leads to the scaled equation
\begin{equation}
  \label{asymptotics3}
 \eps u_T+\eps uu_X = \eps^3 u_{XXT} .% + \delta\eps^2u_{XX}. 
\end{equation}
With the expansion $u(X,T) = \tilde{u}_0(X,T) + \eps \tilde{u}_1(X,T) + \cdots$, we
have the leading order conservation law
$$
\partial_T \tilde{u}_0+ \tilde{u}_0 \partial_X \tilde{u}_0 = 0.
$$
We write the general, implicit solution by characteristics in the form
$$
\tilde{u}_0(X,T)=f(T-\frac{X}{\tilde{u}_0}).
$$
Matching to the inner solution, we have, for $x>0,$ 
$$
\lim_{X\to0+} \tilde{u}_0(X,T)=f(T)=\lim_{\xi\to\infty} u_0(\xi,T)=\frac{A}{1+\half AT}.
$$
The matching for $x<0$ is similar, giving an odd function for the outer solution.
$$
\tilde{u}_0=\frac{A\,\sgn(X)}{1+\half A(T-\frac{X}{\tilde{u}_0})}.
$$
Solving for $\tilde{u}_0$, we find the leading order outer solution 
\begin{equation}
  \label{outer1}
  u \sim u_{out}(X,T) = \frac{A(\half X+\sgn(X))}{1+\half AT}, \quad |X| < A
  T .
\end{equation}
Continuous matching to the constant, far field conditions we obtain 
\begin{equation}
  \label{eq:21}
  u_{out}(X,T) = \sgn(X)A, \quad |X| \ge AT .
\end{equation}

\begin{figure}
  \centering
  \includegraphics[width=0.7\columnwidth]{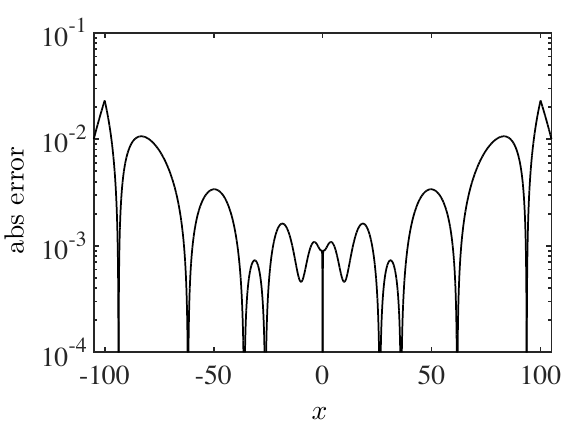}
  \caption{Pointwise error between the uniform asymptotic expansion
    and numerical solution of Fig.~\ref{fig:expansion_shock_sequence}
    at $t = 100$ with $\eps = 0.1$, and $A = 1$.
    \label{fig:expansion_shock_error}}
\end{figure}
\subsection{Uniformly valid asymptotic solution}
Using the standard technique from asymptotics, we can formulate a
composite solution that is asymptotically valid over the entire range
of $x$. Based on the outer solution (\ref{outer1}), (\ref{eq:21}), we
define
$$
F(x,t,\eps)=\left\{\barr{ll} -A, \qquad & x<-At\\[6pt]
\displaystyle\frac{A(\half \eps x+\sgn(\eps x))}{1+\half A\eps t},& |x|<At\\[6pt]
A,  & x>At
\earr
\right.
$$
Then the uniformly valid asymptotic solution is
\begin{equation}\label{uniform}
u(x,t)=\frac{A}{1+\half A\eps t}\left(\tanh \frac x\eps-\sgn(x)\right)+F(x,t,\eps).
\end{equation}

\begin{figure}
  \centering
  \includegraphics[width=\columnwidth]{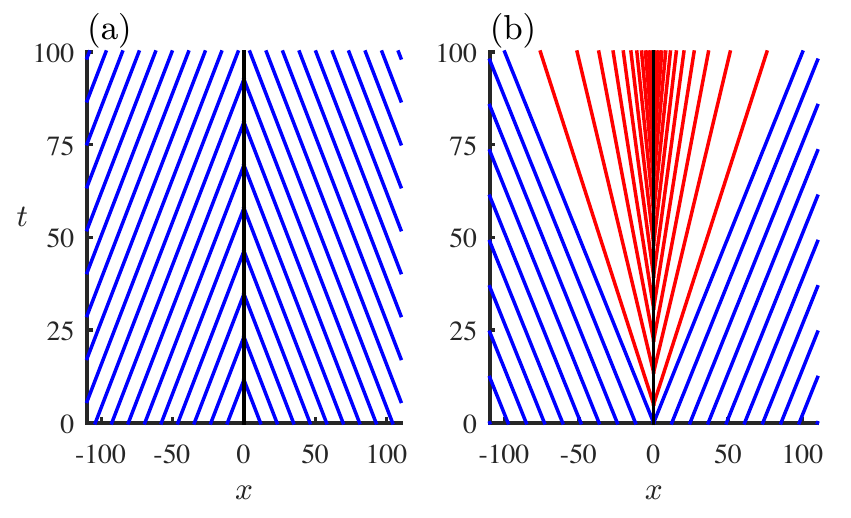}
  \caption{Characteristics for: (a) the compressive shock solution of
    equation (\ref{inviscid_burg}); (b) the expansion 
    shock solution of equation (\ref{eq:1}). 
    \label{fig:characteristics}}
\end{figure}

A comparison of the uniform asymptotic expansion to the numerical
solution is shown in Fig.~\ref{fig:expansion_shock_sequence}.  The two
solutions are hardly distinguishable.  The insets of
Fig.~\ref{fig:expansion_shock_sequence} reveal the smoothed,
non-oscillatory nature of the dispersively regularised expansion
shock.  In contrast, typical dispersive shock waves are characterised
by their oscillatory structure \cite{el_dsws_2016}.  Figure
\ref{fig:expansion_shock_error} displays the absolute error.  Note
that the largest error occurs at the outermost edges of the
rarefaction wave where the asymptotic solution has a weak
discontinuity.  The error in the inner solution is approximately
$\eps^3 = 10^{-3}$, which can be formally identified by going to
higher order terms in eq.~(\ref{asymptotics2}). In
Fig.~\ref{fig:characteristics}(b) we show characteristics calculated
from the outer solution (\ref{outer1}), (\ref{eq:21}) with $\eps =
0.1$ and $A = 1$.

% \FIX{but in eq (\ref{asymptotics2}), it
  % looks like $O(\eps).$ The next non-trivial correction to the inner
  % asymptotic expansion occurs when the $u_T$ term balances the other
  % two.  We could have expanded (\ref{asymptotics1}) as $u = u_0 +
  % \eps^3 u_3 + \cdots$.}

\subsection{Boussinesq equations}

\begin{figure}
  \centering
  \includegraphics[width=\columnwidth]{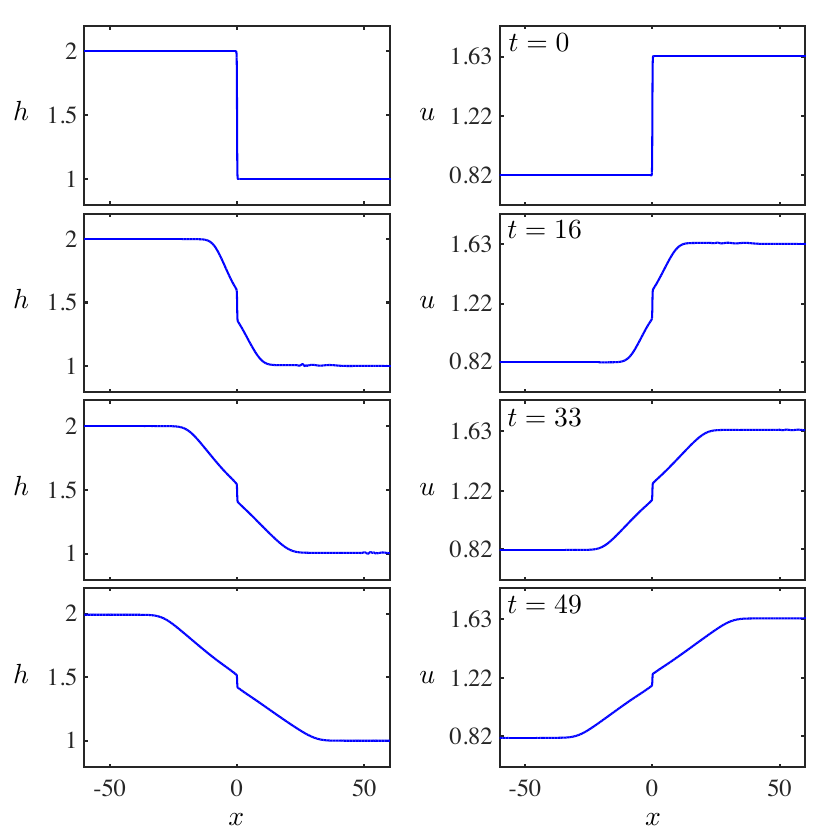} 
  \caption{Evolution of $h$ (left) and $u$ (right) in an expansion
    shock for the Boussinesq system with $h_-=2$, $h_+=1$ and $u_-$,
    $u_+$ given by eq.~(\ref{eq:4}). Jump initial data is smoothed by
    $\tanh(x/\eps),$ as in eq.~(\ref{ic1}), with $\eps = 0.1.$}
     \label{fig:boussinesq_num}
\end{figure}

The Boussinesq equations, formulated in the 1870s \cite{boussinesq},
can take a variety of asymptotically equivalent forms
\cite{whitham_linear_1974}.  While having the same level of accuracy
as the KdV and BBM equations in reproducing dispersive shallow water
dynamics, the Boussinesq equations have the advantage of
bi-directionality.  The system considered here 
\begin{equation}
  \label{boussinesq1} 
  \barr{rcl}
  h_t+(uh)_x&=&0\\[6pt]
  u_t+uu_x + h_x -\third u_{xxt}&=&0,  \earr
\end{equation}
is a reduced, normalised version of an equation that appeared in
\cite{whitham_linear_1974} (see also \cite{bona}), which includes a
$u_x$ term in the dynamical equation for $h$. The non-dimensional
variables $h,u$ represent the height of the water free surface above a
flat horizontal bottom, and the depth-averaged horizontal component of
the water velocity, respectively.  A stationary shock solution of this
system
\begin{equation}
  \label{shockb}
  h(x,t)=\left\{\begin{array}{ll}
      h_-, \quad & x<0\\[6pt]
      h_+, &x>0,
    \end{array}  \right.
  u(x,t)=\left\{\begin{array}{ll}
      u_-, \quad & x<0\\[6pt]
      u_+, &x>0,
    \end{array}
  \right.
\end{equation}
will satisfy Rankine-Hugoniot (RH) jump conditions derived from the
time-independent equations, 
\begin{equation}
  \label{rh2} 
  h_+u_+=h_-u_-; \ \
  h_++\half u_+^2=h_-+\half u_-^2.  
\end{equation}
% For fixed $h_-, u_-,$ the
% first equation is a hyperbola in $h_+, u_+,$ and the second equation
% is a parabola. There are two intersections of the two curves in the
% half plane $h>0,$ one of which is at $(h_,u_-).$ 
The RH conditions (\ref{rh2}) are attained for the two-parameter loci
of states
\begin{equation}
  \label{eq:4}
  u_\pm = h_\mp \left ( \frac{2}{h_- + h_+} \right )^{1/2},
\end{equation}
with arbitrary, positive total water depths $h_\pm$.  In
Fig.~\ref{fig:boussinesq_num}, we show the result of a numerical
simulation demonstrating the persistence of a stationary shock wave
for the Boussinesq system (\ref{boussinesq1}).

The characteristic speeds for the dispersionless system
((\ref{boussinesq1}) with $u_{xxt} \to 0$) are
$\lambda_\pm(h,u)=u\pm\sqrt{h}.$ Therefore, if $u_\pm\geq 0$ as in the
loci (\ref{eq:4}), then the characteristics with speed $\lambda_+$
pass through the stationary shock from left to right. However, the
characteristics with speed $\lambda_-$   leave the shock on both sides if
$u_-<\sqrt{h_-}$ and $u_+>\sqrt{h_+}.$ This is the case for the
choices of $h_\pm, u_\pm$ in Fig.~\ref{fig:boussinesq_num}. These
choices also satisfy the Rankine-Hugoniot conditions for a stationary
shock (\ref{rh2}).  To see that similar data with $u<0$ can make a
stationary shock expansive in the $\lambda_+$ characteristic family,
note that the system (\ref{boussinesq1}) is unchanged under the
transformation $x\to -x$, $u\to -u$.

We remark that the analytical treatment of the Boussinesq expansion
shock appears to be more challenging than it was for BBM.  For
example, there is no clear means to separate variables in an inner
solution due to the nonzero mean values of $h$ and $u$.

\subsection{Discussion}
\label{sec:discussion}

\begin{figure}
  \centering
  \includegraphics[width=\columnwidth]{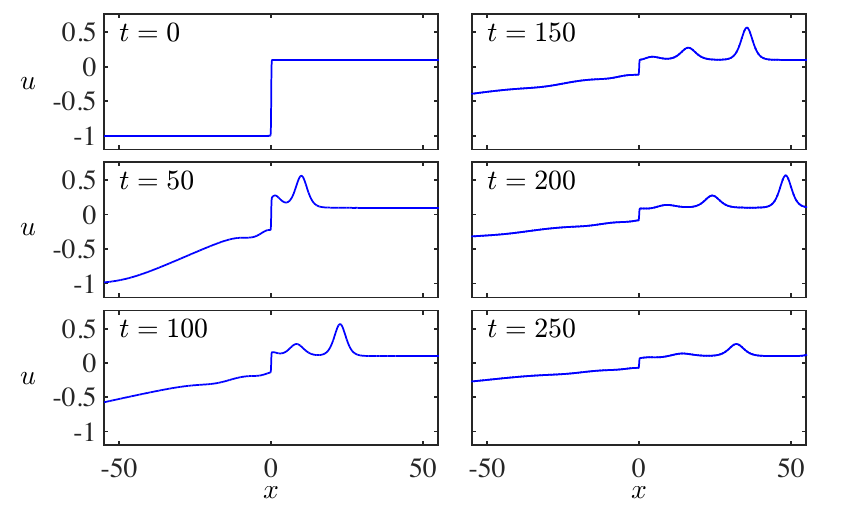}
  \caption{Expansion shock and solitons as components of the initial
    value problem $u(x,0) = 0.55 \mathrm{tanh}(x/\eps)-0.45$ for
    eq.~(\ref{eq:1}) where $\eps = 0.1$.}
  \label{fig:expansion_shock_solitons}
\end{figure}

The expansion shock solutions we have discovered here do decay slowly
in time, but their persistence in the face of the usual rules of
causality is a surprise.  For the BBM expansion shock, we can identify
further robustness to perturbation by considering the asymmetric
initial condition passing through zero
$$
u(x,0)=\half\left((u_+-u_-)\tanh(\frac x\eps) +u_++u_-\right),  
$$
where $ u_-<0<u_+$.  The numerical simulation of eq.~(\ref{eq:1}) with
this asymmetric data is shown in
Fig.~\ref{fig:expansion_shock_solitons}.  As $t$ increases, the
solution quickly develops a stationary, expansion shock with initial
amplitude $A=\min\{u_+,|u_-|\}$, that decays according to the inner
solution (\ref{inner}).  However, the solution also sheds a train of
rank ordered solitons.

The expansion shock also persists in the presence of weak dissipation
in the BBM-Burger's equation
\begin{equation}
  \label{eq:5}
  u_t + uu_x = u_{xxt} + \nu u_{xx},
\end{equation}
where $\nu > 0$ is the dissipation coefficient.  If we consider the
initial data (\ref{ic1}) for (\ref{eq:5}), then the inner solution
exhibits exponential temporal decay
\begin{equation}
  \label{eq:6}
  u_{in}(\xi,T) = \frac{A e^{-\nu T/\eps}}{1+\frac{A \eps}{2 \nu}(1 -
    e^{-\nu T/\eps})} \tanh \xi.
\end{equation}
Thus, for $\nu = \mathcal{O}(\eps)$, the expansion shock decays exponentially as $t\to \infty,$ rather than the 
algebraic decay in the absence of diffusion. In fact, if $\nu \ll \eps$, then (\ref{eq:6}) is
asymptotically equivalent to (\ref{inner}).

The construction presented here can be generalised to higher order
nonlinearity $f(u)$ and higher order, positive differential operators
$\mathcal{L}$ in the form
\begin{equation}
  \label{eq:2}
  \begin{split}
    \mathcal{L}[u_t] + f(u)_x &= \nu u_{xx} ,
  \end{split}
\end{equation}
so long as the non-evolutionary, dispersive character is maintained.
For example, $\mathcal{L} = 1 - \partial_{xx}$ and $f(u) = u^4$ or
$\mathcal{L} = 1 + \partial_{xxxx}$ and $f(u) = u^2$
admit expansion shock solutions that can be approximated with matched
asymptotic methods.

Recalling that the original formulation of the BBM equation
\eqref{eq:1} was as a numerically advantageous shallow water wave
model, \cite{peregrine_calculations_1966}, it is important to stress
that ``engineering'' the dispersion for mathematical or numerical
convenience can lead to new, unintended phenomena, e.g., expansion
shocks.

\subsection{Conclusions}

We have identified decaying expansion shocks as robust solutions to
conservation laws of non-evolutionary type that naturally arise in
shallow water theory.  These models include versions of the well-known
BBM and Boussinesq equations, which are weakly nonlinear models for
uni-directional and bi-directional long wave propagation,
respectively, although as written here, they are not asymptotically
resolved.  The requisite non-local dispersion in these models is not
peculiar to shallow water theory, occurring, for example, in a
Buckley-Leverett equation with dynamic capillary pressure law,
\cite{hassanizadeh1990} describing flow in a porous medium.  Expansion
shocks represent a new class of purely dispersive and
diffusive-dispersive shock waves.  An important open question is
whether expansion shock solutions can be physically realised.\\

This work was supported by the Royal Society International Exchanges
Scheme IE131353 (all authors), NSF CAREER DMS-1255422 (MAH), and NSF
DMS-1517291 (MS).

\appendix
\section{Numerical Method}\label{app}

The numerical methods utilised here for both the BBM (\ref{eq:1}) and
Boussinesq (\ref{boussinesq1}) equations incorporate a standard fourth
order Runge-Kutta timestepper (RK4) and a pseudospectral Fourier
spatial discretisation, similar to the method described in
\cite{el_dispersive_2015}.  We briefly describe the method for BBM
here.

We are interested in solutions $u(x,t)$ to (\ref{eq:1}) that rapidly
decay to the far field boundary conditions $u(\pm \infty,t) = u_\pm$.
The derivative $v = u_x$ therefore rapidly decays to zero and
satisfies
\begin{equation}
  \label{eq:7}
  (1 - \partial_{xx})v_t + (uv)_x = 0 , 
\end{equation}
where $u(x,t) = \int_{-L}^x v(y,t) \,\mathrm{d}y + u_-$.  The Fourier
transform (written $\widehat{f}(k)$ with wavenumber $k$ for a function
$f(x)$) of eq.~(\ref{eq:7}) can therefore be written
\begin{equation}
  \label{eq:8}
  \frac{\mathrm{d}}{\mathrm{d}t} \widehat{v} = - \frac{ik}{1 + k^2}
  \widehat{uv} . 
\end{equation}
The term $\widehat{uv}$ is well-defined because the function $uv$ is
rapidly decaying.  Suitable truncation of the spatial and Fourier
domains turn eq.~(\ref{eq:8}) into a nonlinear system of ordinary
differential equations, which we temporally evolve according to RK4.
The computation of the nonlinear term in (\ref{eq:8}) is efficiently
implemented using the fast Fourier transform (see
\cite{el_dispersive_2015} for further details).  For BBM, $(L,N,\Delta
t) = (200,2^{15},0.01)$ (Figs.~\ref{fig:expansion_shock_sequence},
\ref{fig:expansion_shock_error}), $(300,2^{15},0.01)$
(Fig.~\ref{fig:expansion_shock_solitons}).  For Boussinesq,
$(L,N,\Delta t) = (100,2^{14},0.005)$.

% \enlargethispage{20pt}

% \ethics{This work does not involve any experiments on humans or animals.}

% \dataccess{This paper does not use any experimental data.}

% \aucontribute{All authors contributed equally to problem formulation, solution,
% and manuscript writing.}

% \competing{The authors have no competing interests.}

% \funding{This work was supported by the Royal Society International Exchanges
% Scheme IE131353 (all authors), NSF CAREER DMS-1255422 (MAH), and NSF
% DMS-1517291 (MS)}

\bibliographystyle{vancouver}
\bibliography{exp_shock_r1}

\end{document}